\documentstyle[bezier,aps,preprint]{revtex}
\newcommand{\be}{\begin{equation}}
\newcommand{\ee}{\end{equation}}
\newcommand{\ba}{\begin{eqnarray}}
\newcommand{\ea}{\end{eqnarray}}
\newcommand{\pa}{\partial}

\def\ket#1{|{#1}\rangle }
\def\bea{\begin{eqnarray}}
\def\eea{\end{eqnarray}}

\newtheorem{theo}{\sc Theorem}

\def\U{{\cal U}}

\def\RE{\hbox{R}\hspace{-.20cm}\rule{0.1mm}{2.75mm}\hspace{1.9mm}}

\def\W{{\cal W}}
\def\M{{\cal M}}
\def\V{{\cal V}}
\def\T{{\cal T}}
\def\H{{\cal H}}

\begin{document}
\preprint{DYSCO 056, IFUM 513/FT
 August 1995}
\title{Quantized Temperatures Spectra in Curved Spacetimes}
\author{Marco Bertola, Vittorio Gorini, Mauro Zeni.\cite{emails}}
\address{Dipartimento di Fisica, Universit\`a\ di Milano,
Via Celoria 16, 20133 Milano, II Facolt\`a\ di Scienze Universit\`a\ di
Milano, V. Lucini 3, 22100 Como, I.N.F.N.
- Sezione di Milano, ITALY}
\date{1 August 1995}
\maketitle
\begin{abstract}
We consider the thermal properties of a scalar field theory on
curved spacetimes. In particular, we argue for the
existence in the de Sitter, Kruskal and Rindler  manifolds
of a discrete spectrum of allowed temperatures (the odd multiples of a
fundamental one). For each temperature we give an explicit
construction of
the relative two point function in terms of the lowest temperature one.
These results are actually valid for a wider class of static metrics with
bifurcate Killing horizons, originally studied by Sewell.
Some comments on the interpretation of our results are given.
\end{abstract}
It is well known that  thermal effects occur when quantizing a
field theory on certain kinds of curved spacetimes, the relevant cases being
the Rindler, de Sitter and Kruskal manifolds
\cite{hawk1,hawk2,gibbon,unru}.
In all these cases the presence
of a dimensional parameter in the theory defines a fundamental temperature
$T_0\neq 0$ which characterizes the ground state of the field  when restricted
to a suitable region of the
manifold ({\em wedge}) bounded by horizons.
Inside this region one can identify a  ``time'' variable, associated with  a
suitable time-like Killing field
 and the thermal behaviour displays itself through the properties
of the analytic continuation
of two point functions to {\em complex}
 ``time'' in that the ground state of the field satisfies
the Kubo-Martin-Schwinger (KMS) condition
with respect to the ``time''
evolution at temperatures given by
$T_0={a\hbar\over 2\pi c k}$,
$T_0={\hbar c\over 2\pi R k}$ and
$T_0={\hbar c^3\over 8\pi Mk}$,
for the Rindler, de Sitter and Kruskal cases
respectively. In these formulas
$a$ is the proper acceleration of the Rindler observer, $R$ is the radius
which characterizes the de Sitter manifold and $M$ is the mass of the
eternal Schwarzschild black-hole.
Actually, as shown by Sewell \cite{sewell}, the existence of a fundamental
temperature can be inferred under suitable conditions for a wide class
of static metrics, of which the above ones are special cases.

The question arises as to which thermal equilibrium states of the field
exist at temperatures other than the fundamental one. It is well known that
in the flat minkowskian case and with respect to the time coordinate of an
inertial observer, the fundamental temperature is zero, the ground state
being the usual Minkowski vacu\-um, and that there is a KMS state at any
temperature $T$.\par
In this paper we argue for the fact that the only thermal
states which arise on a manifold of the kind considered and with respect to the
timelike Killing coordinate in the wedge are at temperatures
which are the odd multiples of the fundamental one, and we give an explicit
construction of the corresponding two point functions.
Our results are understood in terms of the compact character of the time
orbits in the complexified extension of the manifolds.\par
We spell out our construction in the case of the de Sitter (dS)
manifold\footnote{In the sequel we use units such that $\hbar=
c=G=k=1$.}.
Such a manifold is the maximally symmetric Lorentz manifold given by the
hyperboloid with equation $\sum_i ( x^{i})^2 -(x^{0})^2 = R^2$ embedded
in the five-dimensional Minkowski ambient space $\RE^5$
from which it inherits
the metric and the isometry group, i.e. the Lorentz group $O(4,1)$.
The world line of a freely falling observer moving
through the point $(0,0,0,0,R)$ and contained in the $(x^{0},x^{4})$-plane
is the geodesic parameterized as
$\gamma=$ $\hspace{-3pt}\{ x^{0}\hspace{-4pt}=R\sinh\frac{t}{R},\;
x^{1}\hspace{-4pt}= x^{2}\hspace{-4pt}=
x^{3}\hspace{-4pt}=0,\;x^{4}\hspace{-4pt}=R\cosh\frac{t}{R}\}$.
The    set of all events of  dS which
can be connected  with the observer by two  light-signals (one future and
one past directed) is called
{\em right wedge} (denoted $\M^+$); it is parametrized
by the coordinate map:
$x(t,{\underline{x}})
=\{x^{0} =  \sqrt{R^2 - r^2 }\sinh \frac{t}{R},\
( x^1,x^2,x^3) = {\underline{x}},\
x^{4}=\sqrt{R^2 - r^2 } \cosh \frac{t}{R}$;
$r\equiv |{\underline{x}}|\hspace{-3pt}<
\hspace{-3pt} R\}$.
The {\em left wedge} ($\M^-$) is defined as above
with the replacement  $x^{4} \to -x^{4}$. The boundary of $\M^+\cup\M^-$
is the union of the null surfaces
$ \H^\pm  = \{x \in dS, \;\;\;x^{0}=\pm x^{4}\}$. These are respectively
the future/past horizon of the geodesic observer. Moreover, call
$\V^{\pm} = \{x \in dS :\,\pm x^0> |x^4| \}$.
To complete the covering
of the manifold requires the additional maps (for $\V^\pm$)
$ x^{0} = \pm \sqrt{ r^2 -R^2 } \cosh(\frac {t} R)$,
$x^{1}$, $x^{2}$, $x^{3}$ as before,
$x^{4} = \sqrt{ r^2 -R^2 } \sinh(\frac {t} R)$, $r>R$.
In terms of the given coordinates the dS metric reads
$ d\tau^2=[1-(\frac r R)^2]dt^2 - [1- (\frac r R)^2]^{-1} dr^2
-r^2(d\theta ^2 +sin^2\theta d\varphi ^2 )$,
where $\theta$ and $\varphi$ are the polar angles of ${\underline{x}}$.
The vector $\pa_t $ is a Killing field which
is timelike in the two wedges and only therein.
Despite the fragmented structure, the $t$ coordinate labels a single
one-parameter isometry group $\T$ of dS,
namely the one induced by the Lorentz
boosts in the $(x^{0},x^{4})$ plane of the ambient space, and for
$s\in {{\RE}}$ one has
\begin{equation}
\T_s [x(t,{\underline{x}})]
=x(t+s,{\underline{x}})\equiv x^{s}.
\label{quattro}
\end{equation}
The orbits of $\T$ are branches of hyperbolas or half lines (on the
Killing horizons). We stress that $\T$ is an isometry defined on the
whole manifold, and moreover it can be defined uniquely
for complex values too as
an isometry $\T_{t+i\sigma}$ in the de
Sitter complexified manifold $dS^{\cal C}$.
Since the parameter $t$  can be interpreted as the proper
time of the observer sitting on the geodesic $\gamma$,
one can call $\T$ the
``time-translation group relative to $\gamma$''
 (see \cite{ugo} for a more detailed
discussion).\par
We now discuss the quantum theory of a (quasi)-free hermitian
bosonic scalar field $\Phi$ on dS.
In this case, the whole content of the theory is
contained in the two point function
$\W(x_1,x_2) = \langle \Omega |\Phi(x_1)\Phi(x_2) \Omega\rangle$.
This kind of theory has been studied repeatedly in the literature. As usual
when one quantizes a field theory on a curved spacetime, there is no natural
prescription for defining the ``vacuum'' state $\Omega$ of the field.
In particular, the requirement of invariance under the full dS isometry group
$O(4,1)$ selects in the massive case
a two--parameter family of possible ``vacua''\cite{allen}.
Different criteria have been given to single out among these a distinguished
state, which has been called
the ``euclidean'' vacuum \cite{gibbon,allen,birrel}.
This state is characterized by the local Hadamard
condition and is non singular on the full Killing horizon \cite{kaywald} (there
is one and only one vacuum of this kind for each mass $m>0$).\par
More recently the special status of this theory has been described in
terms of the analyticity properties of the $\W$'s \cite{ugo}. The
thermal properties of the euclidean vacuum $\ket {\Omega}$
are encoded in the $2\pi iR $
periodic analytic behaviour of the time translated correlation function
$
\langle\Omega | \phi(x_1)\phi(x^t_2)\Omega\rangle=$ $ {\cal W}(x_1,x^t_2)
\equiv $ $\W_{12} (t) $,  and of its permuted
$\langle\Omega | \phi(x^t_2)\phi(x_1)\Omega\rangle=$ $ {\cal W}(x^t_2,x_1)
\equiv $ $ \W '(x_1,x^t_2)\equiv $ $ \W'_{12}(t)$, along the complexified
orbits of $\T$.
Precisely, $\W_{12}(t)$ satisfies the KMS condition at inverse temperature
$\frac 1 {T_0}=\beta_0=2\pi R$ in the sense that, distributionally,
$\W_{12}(t+i 2 \pi R-i0)=\W'_{12}(t-i0)$.
Since the theory is quasifree, the KMS property $\langle\Omega | A B_{t+i
2\pi R-i0}|\Omega\rangle=\langle\Omega |B_{t+i0}A|\Omega\rangle$ extends
immediately to any two fields observables $A$ and $B$ supported in bounded
regions ${\cal O}_1$ and ${\cal O}_2$ contained in the right wedge. The
question as to which thermal states exist at temperatures other than
$T_0$ and defined on the field algebra
on the whole dS manifold, is addressed by the following theorem.
\begin{theo}
Let $\ket{\Omega_\beta}$ be a local
state for a hermitian scalar bosonic quantum field
$\Phi$ on the dS manifold which is KMS in the right wedge
at inverse temperature
$\beta$ w.r.t. the evolution (\ref{quattro}). Then $\beta$ must be
of the form
$\beta_0/(2L+1)$, where $L$ is a nonnegative integer and $\beta_0 =2\pi R$.
\end{theo}

{\bf Proof.}
We express the proof using the smeared field
$\Phi(f) = \int_{dS} dx\sqrt{g(x)} f(x) \Phi(x)$, where $f(\bullet)$ is
an infinitely differentiable real function on dS with compact support.
We consider the smeared two point function
$\W(t)=\int  dx\, dy
\sqrt{g(x)g(y)}f(x) h(y) \langle\Omega_\beta|\Phi(x)\Phi(y^t)|\Omega_\beta
\rangle $
and its permuted $\W'(t)$.
By the KMS condition $\W(t)$ and $\W'(t)$ are
boundary values $\W(t+i0)$ and $\W'(t-i0)$ of functions
 $\W(z)$ and $\W'(z)$  which are holomorphic in the
strips $0< Im\; z<\beta$ and $-\beta< Im\; z< 0$, $Re\; z =t$
respectively,
continuous on the boundary and such that
$\W(t+i\beta -i0)=\W'(t-i0)$.
We prove first that $\beta\leq\beta_0$.
Indeed choose $f$ and $h$ such that their respective supports
${\cal O}_1$ and ${\cal O}_2$ lie in the right wedge and are causally disjoint
(see Fig.1).
Then the microcausality axiom and the hermiticity
of the field imply that there exists a maximal non empty open interval
$I$ , containing the origin, such that
$\W(t)=\W'(t)=\W(t)^*  $ if $t\in I$. Furthermore, $I$ is bounded since
if we let $t$ increase (or decrease) the support of
$h(x^{-t})$ will eventually intersect the causal domain of $f$ and
the function $\W (t)$ will cease to be real (otherwise the fields would
always commute). Since an analytic function on the real axis which is real on
a segment is real on the whole axis, this implies a breaking of analyticity.
Then, because of the periodic structure of the manifold in the
complex $z$ time, the function
$\W (z)$ must have a break of analiticity on the $Im\; z =2\pi
R=\beta_0$ line: indeed, this line must be identified with the real axis on
which $\W(z)$ is not analytical. Then, if we had
$\beta\geq\beta_0$ the function $\W(z)$ would be holomorphic in the
whole strip $0\leq Im \; z\leq\beta$, leading to a contradiction.
Now, since
$\W(z)$ assumes real values with continuity on the segment $I$
of the real axis, by the Schwarz reflection principle we can analytically
continue it across the segment itself by the mere definition
$\W(z)\equiv \W^*(z^*)$, $Im\; z\leq 0$.
Similarly, we can continue $\W(z)$ above the line
$i\beta$ across the ``gate'' $Re\; z\in I$.
Iterating this procedure, we finally
obtain an holomorphic function in the domain
$\U=$ $\{Im\; z\neq n\beta,\ n \hbox{ integer}\}$ $\cup \{ Im\; z=
n\beta,\ Re\; z\in I\}$.
Moreover, since $\W(z)$ takes the same values at every ``gate'', it is
necessarily a periodic function with period $i\beta$ (at $Im\; z =n\beta$
$\W(z)$ is continuous across the gates $Re\; z\in I$, and
discontinuous across the cuts $Re\; z <Inf\; I$, $Re\; z >Sup\; I$).
Besides, due to the angular character of the imaginary part of $z$,
$\W(z)$ must be also $i\beta_0$--periodical.
Then, since $\beta\leq\beta_0$ and since, in particular, $\W(z)$ is
holomorphic in the strips $n\beta <Im\; z <(n+1)\beta$, $n$ positive integer,
it must
be $\frac{\beta_0}\beta =M$, $M$ positive integer.\par
Now note that the function $\W(t+i\frac{\beta_0}2)$ is real for all
values of $t$, since the shift $t\longrightarrow t+i\frac {\beta_0}2$ is a
parity transformation which maps a
wedge in the opposite one, and the two wedges are spacelike separated.
Hence $M$ must be odd, otherwise the line $i\frac{\beta_0}2$ would have to be
identified with the real axis on which $\W(z)$ is not real outside the
gate $I$.\par
Note that the theorem does not assert that such states $\ket{\Omega_\beta}$
exist, but it only rules out all temperatures which are not odd multiples of
$T_0$.
We complete the proof by giving a
construction of the KMS state at inverse
temperature $\beta =\beta_0 /(2L+1)$,
for $L=1,2, \dots $(if it exists), in terms of
the state $\ket{\Omega_{\beta_0}}$ at the fundamental temperature $\beta_0$.
Let $p_1,\; p_2 \in dS$.
Since  $t$ is a Killing coordinate we have
$\W_{\beta}(p_1,p_2)=
\int_{-\infty}^{\infty}d\omega e^{-i\omega (t_1-t_2)}
g_{\beta}(\omega ;\underline{x_1},\underline{x_2} )$
and a similar formula for $\W'_{\beta}$.
Also $g_{\beta}(\omega ) -g'_{\beta}(\omega) = c(\omega )$, independent of
$\beta$, since the commutator is a c-number. In terms of Fourier transforms,
the KMS condition writes $g'_{\beta}(\omega ) = e^{-\beta\omega}g_{\beta}
(\omega )$, so that $g_{\beta}(\omega ) = c(\omega )/(1-e^{-\beta\omega})$
and $g'_{\beta }(\omega ) = c(\omega )/(e^{\beta\omega} -1)$ (in the flat
case we can perform the limit of zero temperature to get
$g(\omega ) = \theta (\omega )c(\omega )$,
$g'(\omega ) = -\theta (-\omega )c(\omega )$).
Now, if $\beta = \beta_0/(2L+1)$, $L$ a positive integer, we have the
identity
\be
\frac 1 {(1-e^{-\beta\omega})} = \frac{\sum_{n=0}^{L} e^{-n\beta\omega}}
{1-e^{-\beta_0\omega}} +
\frac{\sum_{n=1}^{L} e^{n\beta\omega}
}{e^{\beta_0\omega}-1}\quad .\label{sette}
\ee
Then, it follows from the above that $\W_{\beta}$ is given by
\be
\W_{\beta}(p_1,p_2) = \hspace{-4pt}\sum_{n=0}^{L}\hspace{-1pt}
\W_{\beta_0}(\T_{-in\beta}p_1,p_2) +\hspace{-4pt}
\sum_{n=1}^{L}\hspace{-1pt}\W'_{\beta_0}
(\T_{in\beta}p_1,p_2) \ .
\label{otto}
\ee
The minkowskian limit can now be easily performed by letting
$R\rightarrow\infty$ with $\beta$ fixed, namely $L\rightarrow\infty$
and hence $\beta_0 =(2L+1)\beta\rightarrow\infty$, to get the usual formula
for the minkowskian case \cite{birrel}.\par
The higher temperature states $\ket{\Omega_\beta}$ are of course not fully
symmetric
even if  the fundamental state $\ket{\Omega_{\beta_0}}$ is a maximally
symmetric one.
In the flat Minkowskian case the
vacuum state $\ket{\Omega_M}$ is Poincar\'e\ invariant, whereas a thermal
state $\ket{\Omega_{M,\beta}}$ at $T> 0$ is
invariant under space rotations and
spacetime translations but not under a Lorentz boost (Doppler dipole
anisotropy). In the dS case the states $|\Omega_\beta\rangle$ are invariant
under space rotations and of course under the ``time'' translation
(\ref{quattro}), but invariance under those dS transformations which as
$R\longrightarrow \infty$ contract into the group of space translations is
lost. This is easily understandable if we note that as long as $R\neq \infty$
the generators $P_i$ of such transformations do not commute among each
other and their commutator contains the generators of local Lorentz boosts
(geodesic deviation) so that additional invariance under the $P_i$'s would
imply invariance under the full dS group, a property which is enjoyed only
by the fundamental state $|\Omega_{\beta_0}\rangle$.\par
Since the proof of the theorem makes reference
only to the existence of two wedges,
bounded by horizons and causally disconnected, in which $t$ is a Killing
time which is periodic on the imaginary axis in the complex extension of the
manifold, the theorem
extends without modification to the Kruskal \cite{hawk1,hawk2}
and Rindler \cite{unru}
manifolds, which possess the same relevant structure.
Indeed the dS, Schwarzschild and Rindler metrics are all particular cases of
a class of static metrics studied by Sewell \cite{sewell} and for which
the author has given sufficient conditions for the existence of a temperature.
The underlying manifold is of
the form $\M =X\times Y$ (pointwise $x=(\eta ,\xi ;y)$), where $X$ and $Y$ are
two-dimensional and $X$ is an open submanifold of $\RE^2$. The corresponding
Lorentz metric is given by the formula
\be
d\tau^2=A(\eta^2-\xi^2,y)(d\eta^2-d\xi^2)-B(\eta^2-\xi^2,y)d\sigma^2(y)\ ,
\label{quattordici}
\ee
where $A$ and $B$ are positive valued, smooth functions and
$d\sigma^2(y)$ is a positive metric on Y.
We define the following submanifolds of $\M$: $\M^{\pm}=\{\pm\xi >|\eta |\}$,
$\V^{\pm}=\{\pm\eta >|\xi |\}$, $\H^{\pm} =\{\eta =\pm\xi\}$.
$\M^{\pm}$ and $\V^{\pm}$ can be parametrized by coordinates $\rho$ and $s$
as
\be
\eta =\rho \sinh{s},\qquad \xi =\pm\rho \cosh{s}\
\label{quindici}
\ee
for $\M^{\pm}$, and interchanging the definitions of $\eta$ and $\xi$ for
$\V^\pm$.
In terms of such coordinates the metric writes
\be
d\tau^2=\pm A(-\rho^2,y)(\rho^2 ds^2-d\rho^2)-B(-\rho^2,y)d\sigma^2(y)\ ,
\label{diciasette}
\ee
where $+$ refers to $\M^{\pm}$ and $-$ to $\V^{\pm}$. $\M^+$ and $\M^-$
are respectively the right and left wedge and they are bounded by the
bifurcate Killing horizon $\H^+\cup\H^-$. We see from (\ref{diciasette}) that
$s$ is a timelike Killing coordinate in $\M^\pm$.
Moreover, it is seen from (\ref{quindici})
that $\M$ is periodic along the imaginary $s$ axis with period $2\pi$.
Therefore, under these conditions the theorem extends to $\M$ and the
uniformly accelerated particle detector moving along the world line
$\rho =const$, $y=const$ will thermalize in the field at either the
temperature given by the Tolman relation
$
T_0 (\rho ,y)=\frac 1 {2\pi \rho A(-\rho^2, y)^{{1\over 2}}}
\label{diciotto}$,
or at any of the odd multiples of $T_0(\rho ,y)$.
Note the dependence of $T_0$
on $y$. For example, if $y=(\theta ,\varphi )$, $d\sigma^2(y) =d\theta^2 +
\sin^2\theta d\varphi^2 $, $T_0$
depends on the direction as is to be expected
for a static non isotropic metric.
Specializing the above formula to the dS and Schwarzschild cases
one obtains $T_0 (r)= {1\over 2\pi \sqrt{R^2-r^2}}$ for dS
and $T_0 (r)= {1\over 8\pi M\sqrt{1-{2M\over r}}}$ for Schwarzschild.
These temperatures describe an Unruh type effect experienced by a uniformly
accelerated observer in the dS and Schwarzschild
spacetimes, superimposed to the thermal one purely due to curvature.
The dS and
Hawking temperatures ${1\over 2\pi R }$ and ${1\over 8\pi M}$
are obtained in the limit of inertial motion $r=0$ (dS) and $r=\infty$
(Schwarzschild).
In particular, the formula in the dS
case has been  recently established in
\cite{berto_thirring}.\par
Note that our result is based on the assumption that the thermal state be
defined on the whole manifold, thereby being defined even through the
event horizons  and indeed the quantized
temperatures arise from the relation (\ref{quindici})
between the local map and the whole manifold.
In other words, we require
the horizons not to act as physical barriers;
however we strictly do not require local definiteness \cite{HNS,moretti}.
On the other hand, when the
proper analyticity requirements are dispensed with, and the state is considered
only within a single wedge every thermal behaviour may be
displayed and no constraint at all on the temperatures is found.
In  conclusion, quantized temperatures spectra are expected for a quasi--free
field theory on a given Lorentz manifold whenever an imaginary time Killing
orbit is topologically equivalent to a circle (the compactness of the
imaginary time orbit discretizes temperatures in the same way that periodicity
in real time gives a quantization of energy).
Note also that in order for thermal states to arise
it is necessary that the metric be static (in the wedge).
If the metric is stationary but not static, thermal effects are destroyed by
inertial dragging.
For example, Kay and Wald have shown \cite{kaywald} that no
KMS Hadamard state exists for a quasifree field propagating in the Kerr
background.

We conclude with some remarks. Our thermal states represent the static
equilibrium configurations between the fixed background manifold and the
thermal bath of the field quanta; indeed our states correspond to the
Hartle-Hawking  ones \cite{hawk2,kaywald} of the Schwarzschild case.
One may wonder whether
this structured thermal behaviour reflects itself also
in the non equilibrium and
dynamical cases, namely when Unruh states \cite{kaydimock} are considered in
place of equilibrium states or when
the backreaction is properly taken in account and the manifold is free to
fluctuate.
We cannot address this question here, but it seems sensible to believe
that like the Hawking temperature arises both in the static and dynamical
case this should happen to be true for the whole temperature
spectrum as well. If so, and presumably by taking into account the
self--interaction of the field,
the thermal radiation expected, for example from a black
hole, could be spread over different black body spectra; a structured feature
in the radiation emitted from a quantum black hole, although in a quite
different formulation, has been recently pointed out  in
\cite{bekenstein},
where a line emission spectrum is predicted.
One may speculate that also in the dynamical dS case a composite
emission could be relevant
in connection with the cosmological description of the origin of the
cosmic background radiation and the different inflationary scenarios.
\begin{flushleft}
{\bf Acknowledgements}
\end{flushleft}

It is a pleasure to acknowledge stimulating discussions with Jacques Bros and
Ugo Moschella from which the idea of this work has arisen.
M.B. and V. G. also benefitted from the kind hospitality of the
Service de Physique Th\'eorique, CEA--Saclay.
\begin{figure}
\setlength{\unitlength}{0.35cm}
\begin{picture}(11,10)
\thicklines
\scriptsize
\bezier{100}(-.2,0)(4.5,5)(-.2,10)
\bezier{100}(10.2,0)(5.5,5)(10.2,10)
\put(7.85,0){\line(0,1){10}}
\put(2.15,0){\line(0,1){10}}
\put(5,0){\line(0,1){10}}
\put(4,5){\circle{1}}
\put(-1,5){\shortstack{Support\\of $f$\\${\cal O}_1$}}
\put(1.1,5){\vector(1,0){2.7}}
\put(-1,2){\shortstack{Causal\\domain\\of $f$}}
\put(1.1,2){\vector(1,0){2.7}}
\put(1.1,2){\vector(1,3){2.7}}
\put(7,6){\oval(1.4,1)}
\put(6.7,6.5){\vector(0,1){0.6}}
\put(7.3,6.5){\vector(0,1){0.6}}
\put(10,6.7){\shortstack{Action of\\$t$ translation}}
\put(10,6.7){\vector(-1,0){3}}
\put(10,3){\shortstack{Support\\of $h$\\${\cal O}_2$}}
\put(10,3){\vector(-1,1){3}}
\put(4.35,4.7){\line(5,6){4}}
\put(3.5,5){\line(-1,6){1}}
\put(4.35,5.3){\line(5,-6){4}}
\put(3.5,5){\line(-1,-6){0.8}}
\put(5,5){\vector(0,1){1}}
\put(5.1,6){$x^0$}
\put(5,5){\vector(1,0){1}}
\put(5.5,4.4){$x^2$}
\put(14,6.2){\line(0,1){3.7}}
\put(14,2.7){\line(0,-1){2.7}}
\put(15,0){\line(0,1){2.7}}
\put(16,0){\line(0,1){2.7}}
\put(17,0){\line(0,1){2.7}}
\put(15,6.2){\line(0,1){3.7}}
\put(16,6.2){\line(0,1){3.7}}
\put(17,6.2){\line(0,1){3.7}}
\put(18,0){\line(0,1){2.7}}
\put(19,0){\line(0,1){2.7}}
\put(20,0){\line(0,1){2.7}}
\put(18,6.2){\line(0,1){3.7}}
\put(19,6.2){\line(0,1){3.7}}
\put(20,6.2){\line(0,1){3.7}}
\thinlines
\put(7,6.7){\oval(1.4,1)}
\put(7,6.2){\line(1,0){7}}
\put(7,2.7){\line(1,0){7}}
\put(7,3.2){\oval(1.4,1)}
\put(7,5.5){\line(1,0){7}}
\put(14,0){\vector(0,1){10}}
\put(12.5,9.8){$Re\ z$}
\put(19,4.8){$Im\ z$}
\put(14,5.5){\vector(1,0){6}}
\end{picture}
\caption{The ``time'' evolution of the observable $h$ enters the
causal domain of observable $f$.}
\label{tempquan}
\end{figure}
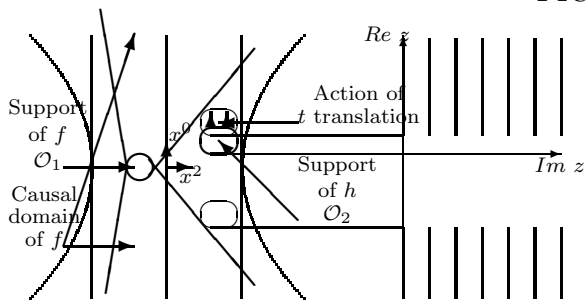

\end{document}